\input harvmac
\input epsf

\def\d{\Lambda}



\Title{}{\vbox{\centerline{The Holographic Dark Energy in a
Non-flat Universe}}}

\centerline{Qing-Guo Huang and Miao Li}
\medskip
\centerline{\it Institute of Theoretical Physics} \centerline{\it
Academia Sinica, P. O. Box 2735} \centerline{\it Beijing 100080}
\medskip
\centerline{\it and}
\medskip
\centerline{\it Interdisciplinary Center of Theoretical Studies}
\centerline{\it Academia Sinica, Beijing 100080, China}
\medskip
\centerline{\it and}
\medskip
\centerline{\it Interdisciplinary
Center for Theoretical Study,} \centerline{\it University of
Science and Technology of China,} \centerline{\it Hefei, Anhui
230026, P. R. China}
\medskip
\centerline{\tt huangqg@itp.ac.cn} \centerline{\tt mli@itp.ac.cn}

\bigskip

We study the model for holographic dark energy in a spatially
closed universe, generalizing the proposal in hep-th/0403127 for a
flat universe. We provide
independent arguments for the choice of the parameter $c=1$ in the
holographic dark energy model. On the one hand, $c$ can
not be less than 1, to avoid violating the second law of thermodynamics.
On the other hand, observation suggests $c$ be very close to 1,
it is hard to justify a small deviation of $c$ from 1, if $c>1$.

\Date{April, 2004}

\nref\agr{A. G. Riess et al., Astron. J. 116 (1998) 1009; S.
Perlmutter et al., APJ 517 (1999) 565. }
\nref\spna{A. G. Riess et al., astro-ph/0402512. }
\nref\bek{J. D. Bekenstein, Phys. Rev. D 7, 2333 (1973); J. D.
Bekenstein, Phys. Rev. D 9, 3292 (1974); J. D. Bekenstein, Phys.
Rev. D 23, 287 (1981); J. D. Bekenstein, Phys. Rev. D 49, 1912
(1994); S. W. Hawking, Phys, Rev. D 13, 191 (1976); G. 't Hooft,
gr-qc/9310026; L. Susskind, hep-th/9409089, J. Math. Phys 36, 6377
(1994). }
\nref\ckn{A. Cohen, D. Kaplan and A. Nelson, Phys. Rev. Lett. 82
(1999) 4971,hep-th/9803132.}
\nref\hsu{S. D. H. Hsu, hep-th/0403052. }
\nref\ml{M. Li, hep-th/0403127. }
\nref\hg{Qing-Guo Huang, Yungui Gong, astro-ph/0403590.}
\nref\gong{Yungui Gong, hep-th/0404030. }
\nref\hor{R. Horvat, astro-ph/0404204.}
\nref\phantom{R. R. Caldwell, M. Kamionkowski and N.N. Weinberg,
astro-ph/0302506, Phys. Rev. Lett. 91 (2003) 071301;
J. M. Cline, S. Jeon and G. D. Moore, hep-ph/0311312.}
\nref\wmap{C. L. Bennett et al., Astrophys. J. Suppl. 148 (2003)
1, astro-ph/0302207; D. N. Spergel, Astrophys. J. Suppl. 148
(2003) 175, astro-ph/0302209. }
\nref\ws{M. Tegmark et al., astro-ph/0310723. }
\nref\bfi{T. Banks and W. Fischler, astro-ph/0307459.}
\nref\ebmore{B. Wang and E. Abdalla, hep-th/0308145;
R. G. Cai, hep-th/0312014, JCAP 0402 (2004) 007;
D. A. Lowe and D. Marolf, hep-th/0402162.}

The time evolution of the cosmic scale factor depends on the
composition of mass-energy in the universe. The cosmological
observations \refs{\agr, \spna} have offered strong evidence that
the expansion of our universe is accelerating, due to dark energy with
negative pressure. Dark energy (or cosmological constant) problem
has been a longstanding problem in fundamental theoretical physics.
It might be fair to say that theorists do not have any clue as to
where this dark energy comes from, and how to compute it from the
first principles. It is therefore a good approach to this problem to
search for plausible phenomenological models, and let experiments
select one of them.

In quantum field theory, we naively estimate the value of the cosmological
constant as the zero point energy with a short distance cut-off scale, for example
the Planck scale, which is exceedingly larger than the observational results.
On the other hand, the cosmological constant problem is a
cosmological problem, since we can ignore the cosmological
constant when we explore the physics at very short distance. We
only need to take the cosmological constant into account when we
investigate the whole universe. Bekenstein et al. \bek\ proposed the
bound $S \leq \pi M^2_p L^2$ on the total entropy $S$ in a volume
$L^3$ and this non-extensive scaling suggests that quantum field
theory breaks down in large scale. A. Cohen et al. \ckn\ proposed
a relationship between UV and IR cut-offs to rescue the local quantum field theory
from the spelling of formation of black holes, this results in a up-bound
on the zero-point energy density. We may view this as the holographic dark
energy, the magnitude of this dark energy is consistent with the cosmological
observations. However, Hsu recently pointed out that the equation of
state is not correct for describing the accelerating expansion of
our universe \hsu. Very recently, one of us \ml\ suggested that
we should use the proper future event horizon of our universe to
cut-off the large scale, resulting in a correct equation of state
of dark energy. The model proposed in \ml\ is consistent with
the Supernova data, which has been discussed in \hg\ in detail, and
has been expanded to the frame of Brans-Dicke scalar-tensor theory in \gong.
A recent discussion on other aspects of holography and dark energy is
\hor.
In this paper, we study this model in the context of a non-flat universe,
extending the work \ml\ to a closed and or an open universe.

As usually believed, an early inflation era leads to a flat universe.
This is not a necessary consequence if the number of e-foldings is not
very large. It is still possible that there is a contribution to the
Friedmann equation from the spatial curvature, though much smaller than
other energy components according to observations. Therefore, it is
not just of academic interest to study a universe with a spatial curvature
marginally allowed by the inflation model as well as observations.

According to \ml, the dark energy density in a flat universe is
\eqn\erd{\rho_\d = 3 c^2 M^2_p R^{-2}_h, }
where we may always let $c > 0$,
$R_h$ is the proper size of the future event horizon,
\eqn\eh{R_h = a(t) \int^{\infty}_{t} {dt' \over a(t')} =
a \int^{\infty}_{a} {da' \over H' a'^2}. }
Since $\Omega_\d = \rho_\d / \rho_{cr}$ and $\rho_{cr} = 3 M^2_p H^2$ is
the critical energy density of our universe,
eq.\erd\ tells us that
\eqn\rhh{H R_h = {c \over \sqrt{\Omega_\d}}. } Taking derivative
with respect to $t$ in both sides of eq.\eh, we obtain
\eqn\deh{\dot R_h = H R_h - 1 = {c \over \sqrt{\Omega_\d}} - 1. }
Following eq.\erd, the changing rate of the holographic dark energy with
time is
\eqn\rrd{{d
\rho_\d \over d t} = -6 c^2 M^2_p R^{-3}_h \dot R_h = - 2 H (1 -
{1 \over c} \sqrt{\Omega_\d}) \rho_\d. }
Because of the conservation of the energy-momentum tensor, the
evolution of the energy density of dark energy is governed by
\eqn\eed{{d \over da} (a^3 \rho_\d) = -3 a^2 p_\d, }
thus we obtain
\eqn\pd{p_\d = - {1 \over 3} {d \rho_\d \over d \ln a} - \rho_\d. }
and the equation of the state of the holographic dark energy is
characterized by the index
\eqn\sthd{w_\d = {p_\d \over \rho_\d} =
 - {1 \over 3} {d \ln \rho_\d \over d \ln a} - 1
= - {1 \over 3} (1 + {2 \over c} \sqrt{\Omega_\d}),  }
where we used $d \ln a = H d t$. The result \sthd\ is a consequence of the
definition of the holographic dark energy, thus is independent of
other components of energy in the universe.
According to this result, we
see that the holographic dark energy has a generic property in a flat
universe: the
index of the equation of state is always  $w_\d
\simeq - {1 \over 3}$ when any other type of
energy dominates, and is always $w_\d = -{1 \over 3} (1 + 2/c)$, when
dark energy dominates. In the latter case, it behaves almost like the cosmological
constant if $c = 1$.

Let us focus on the flat universe first, assuming there is only the holographic
dark energy, $\Omega_\d = 1$.
The entropy of our universe is
\eqn\enp{S = {A \over 4 G} = \pi M^2_p R^2_h. }
If we require the entropy of our universe do not decrease,
we need $\dot R_h \geq 0$ which implies $c \geq 1$ by eq.\deh\
thus $w_\d \geq -1$. The requirement that $c\ge 1$ must be imposed
even when matter and radiation is present, since the universe will be
gradually dominated by dark energy, so $\dot R_h$ approaches
$c-1$ in the far future. Alternatively, if one defines, as motivated
by the AdS/CFT correspondence, the central charge $c\sim M_p^2/H^2$,
we require $\dot H\le 0$, and thus $p\ge -\rho$ (since $\dot H\sim -(\rho
+p)$), again we find $c\ge 1$.

If $c < 1$, we will run into another trouble. The proper size of the
future horizon will shrink to zero, the IR cut-off will become shorter than
the UV cut-off in a  finite time in the future, the very definition of the
holographic dark energy breaks down. $c<1$ corresponds to holographic
dark energy behaving as phantom, here we have provided a global argument
against phantom, which was criticized by local arguments such as instability
and production of gamma rays \phantom.

In the original proposal \ml, $c=1$ is adopted, and the motivation
is the following. If the event horizon is to be regarded as a black hole
horizon, the total energy from dark energy must be determined by the
Schwarzschild relation, this leads to result $c=1$.

Of course it is desirable to have an independent argument for $c=1$.
We have argued that $c\ge 1$. If one can argue
$c\le 1$, then $c=1$ is the only choice. We can not think of a theoretical
argument for $c\le 1$ at the present, the experimental result can be taken
as a support to the choice $c=1$, since, if $c>1$, it can not deviate from
1 too much, and there appears to be no theoretical reason for a small but
non-vanishing deviation. Finally, if $c$ is strictly greater than 1,
in an empty universe when there is only dark energy, the Gibbons-Hawking
entropy \enp\ will increase in time, and it appears rather bizarre that
the empty universe generates entropy constantly. Thus, $c=1$ is a good choice.

In \ml, it was shown that for $c=1$, the cosmic coincidence problem is
resolved if the number of e-foldings is about $60$, the minimal number
required to solve the traditional horizon problem and flatness problem.
The reasoning is simple: for dark energy to be the same order of
the critical density at the present, it must be very small in the end
of inflation. Sufficient inflation red-shifts dark energy quickly,
since $w_\d$ is about $-1/3$ when the inflaton energy dominates.
If $c$ is not equal to 1, this result is still valid, and the resolution
of the cosmic coincidence problem is preserved.

However, one can show that, if the inflaton energy never decays,
then there will be no consistent solution to the Friedmann
equation, since, as can be easily seen from the Friedmann equation,
the holographic dark energy must eventually dominate the unverse.
This is possible only if the inflaton energy decays in a finite time,
so that dark energy is red-shifted to be a very small fraction
in the end of inflation but will dominate the universe in much later time.
An alternative way to see this is the following. If the inflaton energy
never decays and keeps almost as a constant, for {\it all time} the Hubble
constant is determined by the inflaton energy, so the size of the
future horizon is roughly the inverse of the Hubble constant, thus
dark energy can not be red-shifted away. Therefore, a
consistent solution requires the existence of reheating, this is a
rather interesting consequence of the holographic dark energy.
We take this as a support to the holographic dark energy model, since
this rules out an cosmological constant in addition to the holographic
dark energy-after all, all zero-point energy must be included in the
holographic dark energy.
This issue together with other issues will be investigated elsewhere.

Now let us study the case of a universe with a spatial curvature.
As said before, a closed universe with a small positive curvature
($\Omega_k\sim 0.01$) is compatible with observations \refs{\spna, \wmap,
\ws}. We
generalize \ml\ to the spatial closed universe (results about the case of an open
universe with a negative spatial curvature can be obtained from those of
a closed universe by a transformation).
If we still use the definition \erd, when there is only dark energy
and the curvature, $\Omega_\d = 1 + \Omega_k > 1$ and $\rho_\d$ is not
a constant. As a simple example, for $c = 1$, $\dot R_h < 0$.
We know that in a flat universe, the solution with only
dark energy present is a de Sitter space for $c=1$ \ml. We can slice
the de Sitter space with a positively curved spatial section, thus we expect
that in this case dark energy remains a constant.
Thus, we have to modify our definition \erd\ for the holographic dark
energy in a closed universe, one choice is
\eqn\cerd{\rho_\d = 3 c^2 M^2_p L^{-2}, }
here
\eqn\dl{L = a r(t), }
and the definition of $r(t)$ is
\eqn\ddr{
\int^{r(t)}_{0} {dr\over \sqrt{1-kr^2}} = \int^{\infty}_{t} {dt \over a}
= {R_h \over a}, }
or
\eqn\dr{r(t) = {1 \over \sqrt{k}} \sin y, }
where $y = \sqrt{k} R_h / a$. According to eq.\dr, $r(t) = R_h / a$
when $k = 0$ and we get back to the same definition as in a flat universe.
The geometric meaning of $L$ is clear, while $R_h$ is the radial size
of the event horizon measured in the $r$ direction, $L$ is the radius
of the event horizon measured on the sphere of the horizon. We shall see
in a moment that with this definition of the holographic dark energy, the
de Sitter space will be a solution for $c=1$ with only dark energy present.

We want to keep the normalization $a_0=1$, using the energy density
of curvature
\eqn\rk{\rho_k = 3 M^2_p k / a^2, }
we obtain $k = \Omega^0_k H^2_0$.

Using definitions $\Omega_\d = \rho_\d / \rho_{cr}$ and $\rho_{cr} = 3 M^2_p H^2$, we
get
\eqn\chl{HL = {c \over \sqrt{\Omega_\d}}.}
Using \dr, we obtain
\eqn\cdl{\dot L = HL + a \dot r(t) = {c \over \sqrt{\Omega_\d}}
- \sqrt{1-kr^2(t)} = {c \over \sqrt{\Omega_\d}} - \cos y,  }
and
\eqn\cst{w_\d = - {1 \over 3} \left( 1 + {2 \over c} \sqrt{\Omega_\d}
\cos y \right). }
This is a result valid regardless the energy content of the universe.
If we take $c=1$, then $w_\d$ is bounded from below by
\eqn\boundb{w_{\hbox{min}}=-{1\over 3}\left(1+2\sqrt{\Omega_\d}\right).}
Taking $\Omega_\d=0.73$ for the present time, this lower bound is $-0.90$. As we shall
see shortly, $c$ can not be less than $1$, as in the flat case, thus,
$\Omega_\d$ is always bounded by this number. Future experiments will
either confirm this prediction or rule out this dark energy model
based on this number.

According to \cerd\ and \rk, the ratio of the energy density between
curvature and holographic dark energy is
\eqn\rkd{{\Omega_k \over \Omega_\d} = {\rho_k \over \rho_\d}
= {\sin^2 y \over c^2}. }

We now study the behavior of the Gibbons-Hawking entropy in a closed
universe. Since the space is curved in a closed universe, the
Gibbons-Hawking entropy becomes
\eqn\ctp{S = \pi M^2_p L^2. }
Using equation \cdl\ and \ctp,
\eqn\cds{{d S \over dt} = {2 S \over L} {dL \over dt}
= {2 S \over L} \left({c \over \sqrt{\Omega_\d}} - \cos y\right). }
If the entropy can not be decreasing, we must require $\Omega_\d
\cos^2 y \geq c^2$. Applying this condition in \cst, we find that
the holographic dark energy can not be a phantom-like energy
with $w_\d < -1$.
When there is only dark energy and the curvature, $\Omega_\d = 1 + \Omega_k$
and we have
\eqn\cod{\Omega_\d = \left( 1 - {\sin^2 y \over c^2} \right)^{-1}. }
and
\eqn\cdso{{d S \over dt}
= {2 S \over L} \left((c^2 - \sin^2 y)^{1/2} - \cos y\right). }
If $c = 1$, $d S / dt = 0$ and the holographic dark energy behaves like
a cosmological constant with $w_\d = -1$, this is the desired result supporting
the new definition \cerd.
If $c > 1$, $d S / dt > 0$ and
$w_\d > -1$. If $c < 1$, $d S / dt < 0$, the entropy is decreasing with
the evolution with time, and the holographic dark energy looks like
phantom energy with $w_\d < -1$.
So we also require $c \geq 1$ in a closed universe.

To describe our universe, we take matter into account.
We use the
Friedmann equation to relate the curvature of the universe to the
energy density and expansion rate,
\eqn\fde{ \Omega - 1 =
\Omega_k, \quad \Omega= \Omega_m + \Omega_\d= {\rho_m \over
\rho_{cr}} + {\rho_\d \over \rho_{cr}} \quad \hbox{and} \quad
\Omega_k = {\rho_k \over \rho_{cr}}, } where $\rho_m$ ($\rho_\d$)
is the energy density of matter (dark energy). Since
$\rho_m = \rho^0_m a^{-3}$ and $\rho_k = \rho^0_k a^{-2}$, the
energy density of matter and curvature become
\eqn\rmk{\eqalign{\rho_m&=\Omega^0_m \rho^0_{cr} a^{-3}, \cr
\rho_k&=\Omega^0_k \rho^0_{cr} a^{-2}, } } where $\rho^0_{cr} = 3
M^2_p H^2_0$ and we set $a_0=1$, $a = (1+z)^{-1}$, here $z$ is the
cosmological red-shift. Since the energy density of matter
$\rho_m$ decreases faster than $\rho_k$,
$\rho_k$ will become larger than $\rho_m$ in the future, if sufficient
expansion is allowed. Using
\fde\ and \rmk, we obtain \eqn\km{ {\Omega_k \over \Omega_m} = a
\gamma, \quad \hbox{or} \quad 1 - a \gamma = {1 - \Omega_\d \over
\Omega_m}, } where $\gamma = \Omega^0_k / \Omega^0_m < 1$,
constrained by cosmological observations to be no larger than a few
percent. When the curvature density catches up the matter density, $\Omega_k =
\Omega_m$, so $a \gamma = 1$, $\Omega_\d = 1$ or $z = -1
+ \gamma$.

Now using Friedmann equation \fde\ and \rmk, we have
\eqn\ah{{1 \over aH} = {1 \over \sqrt{\Omega^0_m} H_0} \left( {1 - \Omega_\d
\over a^{-1} - \gamma} \right)^{1/2}, }
and
\eqn\rd{\eqalign{\rho_\d &= \Omega_\d \rho_{cr} = {\Omega_\d \over
1 - \Omega_\d} (\Omega_m - \Omega_k) \rho_{cr} = {\Omega_\d \over
1 - \Omega_\d} (\rho_m - \rho_k) \cr &= \rho^0_m {\Omega_\d \over
1 - \Omega_\d} (1 - a \gamma ) a^{-3}. }}
Combining eqs. \cerd, \dr\ and \rd,
\eqn\rhm{L = {a \over \sqrt{k}} \sin y =
a {c \over \sqrt{\Omega^0_m} H_0} \left({1
\over \Omega_\d} {1 - \Omega_\d \over a^{-1} - \gamma}
\right)^{1/2}. }
Substituting eqs. \eh\ and \ah\ into \rhm,
\eqn\inq{{1 \over \sqrt{k}} \sin \left[ {\sqrt{k} \over \sqrt{\Omega^0_m} H_0}
\int^{\infty}_{x} d x' \left( {1 - \Omega'_\d \over
a'^{-1} - \gamma} \right)^{1/2} \right]
= {c \over \sqrt{\Omega^0_m} H_0}
\left( {1 \over \Omega_\d} {1 -
\Omega_\d \over a^{-1} - \gamma} \right)^{1/2}, }
where $x = \ln a$. Taking derivative with respect to $x$ in both sides of
equation \inq, we get
\eqn\dnq{{\Omega'_\d \over \Omega^2_\d} = (1
- \Omega_\d) \left( {2 \over c} {1 \over \sqrt{\Omega_\d}} \cos y
+ {1 \over 1 - a \gamma} {1 \over \Omega_\d} \right), }
where the prime denotes the derivative with respect to $x$.

We pause here to comment that, although the differential equation \dnq\
is a consequence of the integral equation \inq, its solution does not
have to be a solution of the original integral equation, since the integral
equation imposes a boundary condition at $t=\infty$ or $a=\infty$, namely,
$\Omega_\d$ must approach $1$ at $t=\infty$. Thus, the introduction of
the holographic energy always requires that the energy density will be
dominated by dark energy eventually, as the existence of event horizon
indicates.

Applying eqs. \cst\ and \dnq, we derive the rate of evolution of $w_\d$
with the cosmological red-shift $z$
\eqn\rmd{{d w_\d \over d z} =
{1 \over 3c} \sqrt{\Omega_\d} \left[ {1 - \Omega_\d \over
1 - \gamma (1 + z)^{-1}}
+ {2 \over c} \sqrt{\Omega_\d} (1 - \Omega_\d \cos^2 y) \right]
{1 \over 1 + z}. }
Since $w_k = \rho_k / p_k = -1 / 3$,
\eqn\acc{{\ddot a \over a} = - {1 \over 6 M^2_p} (\rho_\d + 3 p_\d
+ \rho_m) = {1 \over 6 M^2_p} \left( {2 \over c} \sqrt{\Omega_\d} \cos y -
{\rho_m \over \rho_\d} \right) \rho_\d. }
Once the expansion of our universe starts to  accelerate, it
will accelerate forever.

When our universe is dominated by matter, $\Omega_\d \ll 1$, we
get $w_\d \simeq - 1/3$ and $\rho_\d \propto a^{-2}$. On the
other hand, after of the time when $\Omega_m = \Omega_k$ or $z =
-1 + \gamma$, $1 - a \gamma$ will be always negative and
$\Omega_\d$ becomes always larger than one. According to eq. \dnq,
$\Omega'_\d$ is positive and $\Omega_\d$ increases during
the period of $\Omega_\d < 1$.
After the point
$\Omega_\d=1$, $\Omega_\d$ will increase for a while until reaching
a turning point. Passing over the turning point, $\Omega'_\d$
will become negative, and $\Omega_\d$ will steadily decrease to 1.
The condition for $\Omega'_\d < 0$ is
\eqn\tn{{2 \over c} \sqrt{\Omega_\d} \cos y + {1 \over 1 - a
\gamma} > 0. }
Solving this relation, we obtain
\eqn\tnz{z < z_m = -1 +
\gamma \left( 1 + {c \over 2 \sqrt{\Omega_\d}} {1\over \cos y} \right)^{-1}. }
It says that $\Omega_\d$ will decrease after $z = z_m$,
but $\Omega_\d$ is still larger than one.
According to equation \ah, $dt / da = 1/(aH)$ is
always regular and we need a finite time to reach $z = z_m$.
The density parameter of the holographic dark energy
$\Omega_\d$ reaches its maximum value at $z = z_m$.
Given the observational result for $(\gamma\le 0.1)$, for a reasonable
choice of $c$ (certainly for $c=1$), the turning point $1+z<1$, thus lies in
future.
We also notice that the quantity $\sqrt{\Omega_\d} \cos y$
does not reach its maximum value when $z = z_m$.
After a straightforward calculation, we find the value of
$(\sqrt{\Omega_\d} \cos y)$ reaches its maximum value at
\eqn\dym{z = z_M =  -1 +
\gamma \left( 1 + {c \over 2 \sqrt{\Omega_\d}}
{1 - \Omega_\d \over 1 - \Omega_\d \cos^2 y} \right)^{-1}. }

Combining eqs. \rkd, \fde\ and \km, we get
\eqn\cdm{\Omega_\d = \left[ 1 - \left( 1 - {1 \over a \gamma} \right)
{\sin^2 y \over c^2} \right]^{-1}. }
Thus we obtain
\eqn\cy{{c \over \sqrt{\Omega_\d}} - \cos y
= \left(c^2 - \sin^2 y + {1 \over a \gamma} \sin^2 y \right)^{1/2} - \cos y. }
According to this equation, if $c \geq 1$, the index of the state of
the holographic dark energy will be always larger than $-1$.
According to \cst, the holographic dark energy behaves as the
phantom-like energy with $w_\d < -1$
if the the maximum value of $\sqrt{\Omega_\d} \cos y$
is larger than $c$, for $c < 1$.

It is interesting to explore the fate of this spatial closed
universe. In the far future $z \rightarrow -1$ and $a \rightarrow
\infty$, the energy density of matter and curvature will be
red-shifted to be exceedingly smaller than dark energy, dark
energy will dominate our universe and ${\Omega_\d} \rightarrow
{1^+}$. Using equation \rkd, we find \eqn\ffy{\cos y = \sqrt{1 -
\sin^2 y} = \sqrt{1 - c^2 {\Omega_k \over \Omega_\d}} \rightarrow
1. } Now equation \dnq\ becomes \eqn\ff{{d \Omega_\d \over d \ln
a} = {2 \over c} (1 - \Omega_\d). } Solving this equation, we
obtain \eqn\ffm{\Omega_\d \sim 1 + \sigma a^{-2 / c}, } where
$\sigma$ is a integration constant and $\sigma a^{-2} \ll 1$. This
result is consistent with our above result $\Omega_\d \rightarrow
1^+$, and guarantees that a solution to the differential equation
\dnq\ is also a solution to the integral equation \inq. Using
equation \cst, we obtain
\eqn\fmd{w_\d \sim -1 - {1 \over 3}
\sigma a^{-2 / c}, } and the evolution of the energy density of
dark energy is \eqn\frd{\rho_\d \simeq {\Omega_\d \over \Omega_\d
- 1} \rho_k \simeq {1 \over \sigma a^{-2}} \rho^0_k a^{-2} =
\sigma^{-1} \rho^0_k a^{-2 (1-1/c)}.  } In order that the energy
density of the holographic dark energy will always be finite, we
require $c \geq 1$ which is consistent with our above discussion.
Specially the energy density of the holographic dark energy will
become a constant for the case with $c = 1$. If $c > 1$, the
energy density of the holographic dark energy will be also
red-shifted to zero asymptotically in the far future.

Setting $k = 0$, we find  $- {1 \over 3}
(1 + 2/c) \leq w_\d \leq - {1 \over 3}$ in the flat case, the same as the results in
\ml. For the open universe, we obtain all the relevant results from
those presented above by performing the simple
transformation: $k \rightarrow - k$, $\rho_k \rightarrow -
\rho_k$, $\Omega_k \rightarrow - \Omega_k$ and $\gamma \rightarrow
- \gamma$. When $z = -1 + \gamma$, the energy density of matter
equals the energy density of curvature and $1 - \Omega_\d = 2
\Omega_m = 2 \Omega_k$. In this case $\Omega_\d$ is always
smaller than one.
The expansion of our universe will also never decelerate.

To summarize, in this paper we generalized the holographic
dark energy in \ml\ to a non-flat universe, and studied the evolution of this dark energy
in the spatial closed universe in detail. In particular, we argued that
in a flat universe as well as in a closed universe, to preserve the
second law of thermodynamics, the parameter $c$ must be no less than 1, thus
it is impossible to have a phantom-like holographic dark energy.

The present approach to dark energy relates the cosmic coincidence problem
to the minimal number of e-foldings \ml, this resolution is quite similar to the
up-bound on the number of e-foldings derived by Banks and Fischler \bfi\ (for
related discussions, see \ebmore), this is not surprising, since the holographic
dark energy takes the energy bound with a horizon into account, while Banks-Fischler's
argument uses the entropy bound with a horizon, the latter is not as tight
as the former.

In two aspects the holographic dark energy model is superior to other models of
dark energy, for instance quintessence models. For one thing, our model is
motivated by holography which is expected to hold in a quantum gravity theory,
thus the model is almost unique, as we have argued that the only parameter
$c$ is to be set to 1, and there is no room for a cosmological constant in addition
to dark energy. For another, models such as quintessence models do not really
solve the cosmic coincidence problem, since we need to set a new energy scale in the
end of inflation, though this scale is not as low as the current Hubble scale.
In our model, we relate the present dark energy density to the minimal number
of e-foldings. Of course one may question why we should choose this minimal
number. One answer to this question is an application of a rather weak
anthropic principle, the number of e-foldings can be arbitrary, but the choice
$60$ is rather generic. Another answer to this question is that after all
this number is needed to solve the traditional cosmological naturalness problems,
we only need one solution to all these problems. In the end, Nature may introduce
only one input, the eventual size of the event horizon, all other parameters
such as number of e-foldings and the CMB power spectrum are all consequences
of this input.

\bigskip
Acknowledgments.

This work was supported by a ``Hundred People Project" grant of
Academia Sinica and an outstanding young investigator award of NSF
of China. This work was initiated during a visit to the
Interdisciplinary Center for Theoretical Study at University of
Science and Technology of China.

\listrefs
\end